\documentclass[superscriptaddress,amsmath,amssymb,twocolumn,11pt,]{revtex4}

\usepackage{amsthm}
\usepackage{charter}
\usepackage{graphicx}
\usepackage{bm}
\usepackage{dsfont}
\usepackage[landscape,papersize={297.1mm,210mm},left=1.9cm,right=1.6cm,top=2.7cm,bottom=2.8cm]{geometry}
\usepackage[pdfstartview=FitH, colorlinks, pdfborder=001, linkcolor=red, anchorcolor=blue, citecolor=blue, urlcolor=blue]{hyperref}

\newcommand{\bra}[1]{{\left\langle #1 \right|}}
\newcommand{\ket}[1]{{\left| #1 \right\rangle}}
\begin{document}
\title{Unified monogamy relations for the generalized $W$-class states beyond qubits}

\author{Zhong-Xi Shen}
\email{18738951378@163.com}
\affiliation{School of Mathematical Sciences, Capital Normal University, Beijing 100048, China}
\author{Wen Zhou}
\email{2230501027@cnu.edu.cn}
\affiliation{School of Mathematical Sciences, Capital Normal University, Beijing 100048, China}
\author{Dong-Ping Xuan}
\email{2230501014@cnu.edu.cn}
\affiliation{School of Mathematical Sciences, Capital Normal University, Beijing 100048, China}
\author{Zhi-Xi Wang}
\email{wangzhx@cnu.edu.cn}
\affiliation{School of Mathematical Sciences, Capital Normal University, Beijing 100048, China}
\author{Shao-Ming Fei}
\email{feishm@cnu.edu.cn}
\affiliation{School of Mathematical Sciences, Capital Normal University, Beijing 100048, China}

\begin{abstract}
The monogamy of entanglement stands as an indispensable feature within multipartite quantum systems. We study monogamy relations with respect to any partitions for the generalized $W$-class (GW) states based on the unified-($q,s$) entanglement (UE). We provide the monogamy relation based on the squared UE for a reduced density matrix of a qudit GW state, as well as tighter monogamy relations based on the $\alpha$th ($\alpha\geq2$) power of UE. Furthermore, for an $n$-qudit system $ABC_1...C_{n-2}$, generalized monogamy relation and upper bound satisfied by the $\beta$th ($0\leq\beta\leq1$) power of UE for the GW states under the partition $AB$ and $C_1...C_{n-2}$ are established. In particular, two partition-dependent residual entanglements for the GW states are analyzed in detail.

\medskip
\noindent Keywords: Monogamy of entanglement, Unified-($q,s$) entanglement, Generalized $W$-class state
\end{abstract}
\parskip=3pt
\maketitle

\section{Introduction}
Entanglement is an remarkable phenomenon in quantum mechanics, serving as a vital resource for quantum information processing and communication~\cite{JM2017,MYW2018,HHL2018,DFG2017}. Significant progresses have been achieved in understanding the roles played by the entanglement in quantum tasks such as quantum  teleportation~\cite{BCH1895}, quantum key  distribution~\cite{BM2009} and quantum computing~\cite{RR2001}.

The monogamy of entanglement (MoE) is a distinguishing feature of entanglement that sets it apart from classical correlations~\cite{CKW2000,BMT2004}. A subsystem entangled with one party cannot freely share its entanglement with other parties of the whole system. Since MoE imposes limitations on the potential information accessible to an eavesdropper for extracting the secret key, it holds immense significance in securing various information-theoretic protocols like quantum key distribution~\cite{MP2010,AAC2006,MT2013}. MoE has been also widely studied in many areas of quantum physics such as quantum information theory~\cite{MPS2010}, condensed-matter physics~\cite{XM2011} and even black-hole physics~\cite{EV213}.

MoE manifests itself quantitatively in the form of mathematical inequalities. Coffman, Kundu and Wootters~(CKW) first characterized the monogamy of an entanglement measure $\mathcal{E}$ for a three-qubit states $\rho_{ABC}$~\cite{CKW2000}, known as the CKW inequality,
\begin{equation}\label{CKW}
\mathcal{E}(\rho_{A|BC})\geq \mathcal{E}(\rho_{AB})+\mathcal{E}(\rho_{AC}),
\end{equation}
where $\rho_{AB}={\rm tr}_C(\rho_{ABC})$, $\rho_{AC}={\rm tr}_B(\rho_{ABC})$ are the reduced density matrices of $\rho_{ABC}$, $\mathcal{E}(\rho_{A|BC})$ stands for the entanglement under the bipartition $A$ and $BC$. Subsequently, Osborne and Verstraete further extend the monogamy inequality by incorporating the squared concurrence for any $n$-qubit systems~\cite{T.J.Osborne}. Since then considerable researches have been conducted on MoE by focusing on various entanglement measures such as the squared entanglement of formation (EoF)~\cite{Oliveira2014,Bai3,Bai2014}, the squared R\'{e}nyi-$\alpha$ entropy~\cite{R2015}, the squared Tsallis-$q$ entropy~\cite{Luo2016} and the squared Unified-$(r, s)$ entropy~\cite{KH2019}. Studies on monogamy inequalities based on the $\alpha$th power of entanglement measures for multiqubit systems have been given in \cite{Zhu2014,Luo2015,Luo2016}. The traditional monogamy inequality (\ref{CKW}) provides a lower bound for ``one-to-group'' entanglement, termed as quantum marginal entanglements~\cite{Walter2013}.

Generally, the relation (\ref{CKW}) may not hold for multi-qudit systems. In \cite{YCO2007,JB2008,Bai2014} the authors presented counterexamples in higher-dimensional systems. In 2016 Lancien et al. demonstrated the existence of multipartite higher-dimensional systems in which any non-trivial monogamy relations are not satisfied, based on a class of additive entanglement measures~\cite{CL2016}. Up to now, it appears that only one known entanglement measure, the squashed entanglement, is monogamous for arbitrary dimensional systems~\cite{Christandl2004}. Due to the importance of the study on monogamy relations for higher-dimensional multipartite systems, it is natural to explore the monogamy inequalities for higher-dimensional multipartite states.

With respect to higher-dimensional quantum states, in 2008 Kim and Sanders extended the  GW state from $n$-qubit to $n$-qudit systems, demonstrating that the GW states adhere to the monogamy inequality in terms of squared concurrence~\cite{JB2008}. In 2015 Choi and Kim showed that a superposition of the generalized $W$-class states and the vacuum (GWV) states satisfy the strong monogamy inequality in terms of the squared convex roof extended negativity~\cite{JHC2015}. In 2016 Kim showed that a partially coherent superposition of a GW state and the vacuum saturates the strong monogamy inequality~\cite{JJ2016}. In 2020 Shi and Chen presented the monogamy inequalities beyond qubits by using the Tsallis-$q$ entanglement for the GW states~\cite{SX2020}. Then, Liang et al. adopted a similar methodology to extend the monogamy relations from Tsallis-$q$ entanglement to R\'{e}nyi-$\alpha$ entanglement for the GWV states~\cite{LYY2020}. Furthermore, Li et al. have recently introduced monogamy relations for multi-qudit GW states by using the unified-($q,s$) entanglement \cite{LB2024}. Motivated by these significant advancements, our research aims to further investigate MoE for GW states with arbitrary partitions in higher-dimensional quantum systems.

The unified-($q,s$) entanglement is a two-parametric generalization of the entanglement of formation. For selective choices of the two parameters $q$ and $s$, other entanglement measures such as concurrence, entanglement of formation, Tsallis-$q$ entanglement and R\'{e}nyi-$\alpha$ entanglement can be regarded as the special cases of the unified-($q,s$) entanglement~\cite{JB2011}. It has been proved that the unified-($q,s$) entanglement satisfies the CKW inequality (\ref{CKW}) and its dual inequality~\cite{JB2011,JJ2012}. Khan et al. presented the monogamy relation based on the squared unified-($q,s$) for arbitrary multi-qubit mixed states in the extended ($q,s$) region~\cite{KH2019}. In Ref.\cite{Yang2022} the authors provided universal entanglement distribution inequalities for multipartite higher-dimensional pure states by utilizing the unified-($q,s$) entanglement.

In Ref.\cite{MZX2010}, the authors introduced two partition-dependent residual entanglements (PREs) based on the negativity for several typical multi-qubit states. Furthermore, the authors demonstrated the unique utility of PREs in analyzing the entanglement dynamics of multi-qubit systems, particularly in processing qubit blocks and sub-blocks. PREs can facilitate a comprehensive comprehension of the entanglement dynamics exhibited by GW states with different levels and formats of partitions.

There have been ample quantitative researches and characterizations of restricted shareability for multi-qubit entanglement. However, the understanding of entanglement distribution in higher-dimensional systems is still limited and requires further investigation. In this paper, we consider the monogamy relations of the unified-($q,s$) entanglement (UE) in higher-dimensional systems. This article is organized as follows. In Section \ref{SEC2}, we briefly introduce a few definitions of entanglement measures, as well as an overview of the GW states. In Section \ref{SEC3}, we first provide the extended ($q,s$) region of the generalized analytic formula of UE. By using the analytical formula, the monogamy relation based on the squared UE for qudit GW states is presented. In Section \ref{SEC4}, in order to provide a more precise description of the entanglement distribution of GW states, we delve into tighter monogamy relations based on the $\alpha$th ($\alpha \geq 2$) power of UE. In Section \ref{SEC5}, for $n$-qudit systems $ABC_1...C_{N-2}$, generalized monogamy relation and upper bound satisfied by the $\beta$th ($0\leq\beta\leq1$) power of UE for the GW states under the partition $AB$ and $C_1...C_{N-2}$ are established. Moreover, in Section \ref{SEC6}, we explore the applications of our results in PREs, and offer valuable insights into the study of entanglement dynamics for GW states. Finally conclusion is made in Section \ref{SEC7}.

\section{Preliminaries}\label{SEC2}
We recall some relevant entanglement measures and introduce the concept of GW states.
Let $H_{A}$ and $H_{B}$ denote a finite dimensional complex inner product vector space associated with quantum subsystems $A$ and $B$, respectively.
For a bipartite pure state $|\psi\rangle_{AB}\in H_{A}\otimes H_{B},$ the concurrence  $C(|\psi\rangle_{AB})$ is defined by \cite{PR2001}
\begin{equation}\label{C}
C(|\psi\rangle_{AB}) = \sqrt{2[1-{\rm tr}(\rho_{A}^{2})]},
\end{equation}
where $\rho_{A} = {\rm tr}_{B}(|\psi\rangle_{AB}\langle\psi|)$ is the reduced
density matrix of subsystem $A$.
For any mixed state $\rho_{AB}\in H_{A}\otimes H_{B}, $ the concurrence is given via the  convex roof extension
\begin{equation}\label{CC}
C(\rho_{AB}) = \min_{\{p_{i}, |\psi_{i}\rangle\}}\sum_{i}p_{i}C(|\psi_{i}\rangle),
\end{equation}
where the minimum is taken over all possible pure decompositions of $\rho _{AB}= \sum_{i} {p_i | \psi _i \rangle _{AB} \langle \psi _i |}$.

For a bipartite pure state $|\psi\rangle_{AB}\in H_{A}\otimes H_{B},$  the unified-($q,s$) entanglement is given by
\begin{equation}\label{UE}
U_{q,s}\left(|\psi\rangle_{AB} \right)=U_{q,s}(\rho_A)=\frac{1}{(1-q)s}\left[{\left({\rm tr} \rho^{q}\right)}^s-1\right].
\end{equation}
For a mixed state $\rho_{AB}$, the unified-$(q,s)$ entanglement is given via the convex-roof extension~\cite{JB2011},
\begin{equation}\label{UEC}
U_{q,s}\left(\rho_{AB} \right):=\min \sum_i p_i U_{q,s}(\ket{\psi_i}_{AB}),
\end{equation}
with the minimum taking over all possible pure state
decompositions of $\rho_{AB}=\sum_{i}p_i\ket{\psi_i}_{AB}\bra{\psi_i}$.

As the unified-$(q,s)$ entropy converges to the R\'enyi-$\alpha$ and Tsallis-$q$
entropies when $s$ tends to 0 and 1, respectively, one has
\begin{eqnarray}
\lim_{s\rightarrow 0}U_{q,s}\left(\rho_{AB} \right)={\mathcal R}_{\alpha}\left(\rho_{AB} \right),
\end{eqnarray}
where ${\mathcal R}_{\alpha}\left(\rho_{AB} \right)$ is the R\'enyi-$\alpha$ ($\alpha=q$)
entanglement of $\rho_{AB}$, and
\begin{eqnarray}
\lim_{s\rightarrow 1}U_{q,s}\left(\rho_{AB} \right)={\mathcal T}_{q}\left(\rho_{AB} \right),
\end{eqnarray}
where ${\mathcal T}_{q}\left(\rho_{AB} \right)$ is the Tsallis-$q$
entanglement of $\rho_{AB}$. When $q$ tends to 1,
\begin{eqnarray}
\lim_{q\rightarrow1}U_{q,s}\left(\rho_{AB} \right)=E_{ f}\left(\rho_{AB} \right),
\end{eqnarray}
where $E_{ f}(\rho_{AB})$ is the EoF of $\rho_{AB}$.
Thus unified-$(q,s)$ entanglement is a two-parameter generalization of EoF.

Moreover, for a bipartite pure state $|\psi\rangle_{AB}$ with
Schmidt-rank 2, we have
\begin{eqnarray}
U_{\frac{1}{2},2}\left(\rho_{AB} \right)=C\left(\rho_{AB} \right)
\end{eqnarray}
for $q =\frac{1}{2}$ and $s=2$, and
\begin{eqnarray}
U_{2,1}\left(\rho_{AB} \right)=\frac{1}{2}C^{2}\left(\rho_{AB} \right)
\end{eqnarray}
for $q =2$ and $s=1$.

For any two-qubit mixed state and any bipartite pure state with Schmidt-rank 2, one has \cite{JJ2012}
\begin{equation}\label{D1}
U_{q,s}\left(\ket{\psi}_{AB}
\right)=f_{q,s}\left(C(\ket \psi_{AB}) \right),
\end{equation}
where $f_{q,s}(x)$ is a differential function,
\footnotesize\begin{eqnarray}\label{H1}
f_{q,s}(x)=\frac{\left(\left(1+\sqrt{1-x^2}\right)^{q}
+\left(1-\sqrt{1-x^2}\right)^{q}\right)^s-2^{qs}}{(1-q)s2^{qs}},
\end{eqnarray}\normalsize
with $q\geq1$, $0 \leq s \leq1$ and $qs \leq 3$.

Based on an extended ($q,s$)-region with $q\geq(\sqrt{9s^{2}-24s+28}-(2+3s))/(2(2-3s))$, $0\leq s\leq1$, $qs\leq(5+\sqrt{13})/2$, the authors in Ref.\cite{KH2019} proved that the analytic formula (\ref{D1}) of the unifed-($q,s$) entanglement still holds in parameter region
\footnotesize$$
\mathcal{R}=\left\{
 (q,s)
\;\Bigg|\;\begin{array}{ll}\frac{\sqrt{9s^{2}-24s+28}-(2+3s)}{2(2-3s)}\leq q,\\ q\leq (5+\sqrt{13})/2s,~0\leq s\leq1
\end{array}
\right\}.
$$\normalsize

The $n$-qubit $W$-class states and $n$-qudit GW states are given by \cite{JB2008}
\begin{eqnarray}\begin{aligned}\label{W}
&\ket{W}_{A_1 A_2 ... A_n}\\  &=
a_1 \ket{10\cdots0}+a_2 \ket{01\cdots0}+...+a_n \ket{00\cdots1}
\end{aligned}\end{eqnarray}
and
\small\begin{eqnarray}\begin{aligned}
&\left|W_n^d \right\rangle_{A_1\cdots A_n}\\&=\sum_{i=1}^{d-1}(a_{1i}{\ket {i0\cdots 0}} +a_{2i}{\ket {0i\cdots 0}}+\cdots +a_{ni}{\ket {00\cdots 0i}}),
\label{GW}
\end{aligned}\end{eqnarray}\normalsize
with the normalization conditions $\sum_{i=1}^{n}|a_i|^2 =1$ and $\sum_{s=1}^{n}\sum_{i=1}^{d-1}|a_{si}|^2=1$, respectively.
(\ref{GW}) includes $n$-qubit $W$-class
states in Eq. (\ref{W}) as a special case of $d=2$. The GW state can be viewed as a special case of the
coherent superposition of a generalized W-class state and vacuum (GWV) state,
\begin{align}
\ket{\varphi}_{A_1 A_2 ... A_n}=\sqrt{p}\left|W_n^d \right\rangle_{A_1\cdots A_n}+\sqrt{1-p}
\left|0\right\rangle^{\otimes n}_{A_1\cdots A_n},
\label{GWV}
\end{align}
where $0\leq p\leq1$.

\section{ Monogamy of the unified-($q,s$) entanglement}\label{SEC3}
Consider an $n$-qudit GW state $\left|W_n^d \right\rangle_{A_1\cdots A_n}$ given in (\ref{GW}). We first present a functional relation between UE and concurrence, from which we derive the related monogamy relations. We need the following lemmas.

\noindent{[\bf Lemma 1]}. \cite{KH2019}
The function $f_{q,s}(C)$ with $(q,s)\in\mathcal{R}$ is a monotonically increasing and convex function of concurrence $C$.

Set $y^{2}=x$ and denote $g_{q,s}(y^{2})=f_{q,s}(x)$. Then for any two-qubit mixed states and any bipartite pure states with Schmidt-rank 2, Eqs. (\ref{D1}) and (\ref{H1}) can be rephrased as
\begin{equation}\label{D2}
U_{q,s}\left(\ket{\psi}_{AB}
\right)=g_{q,s}\left(C^{2}(\ket \psi_{AB}) \right),
\end{equation}
where $(q,s)\in\mathcal{R}$  and the function $g_{q,s}(y)$ has the form
\footnotesize\begin{eqnarray}\label{H2}
g_{q,s}(y):=\frac{\left(\left(1+\sqrt{1-y}\right)^{q}
+\left(1-\sqrt{1-y}\right)^{q}\right)^s-2^{qs}}{(1-q)s2^{qs}}.
\end{eqnarray}\normalsize

\noindent{[\bf Lemma 2]}.~\cite{KH2019} The function $g_{q,s}^{2}(C^{2})$  with $(q,s)\in\mathcal{R}$ is a monotonically increasing and convex function of the squared concurrence $C^{2}$.

\noindent{[\bf Lemma 3]}. \cite{JHC2015}
Let $\rho_{A_{j_1}A_{j_2}\cdots A_{j_{m}}}$ be an $m$-qudit reduced density matrix of the $n$-qudit GWV state (\ref{GWV}) $\ket{\varphi}_{A_1\cdots A_n}$, $2 \leq m \leq  n-1$.
For any pure state decomposition of $\rho_{A_{j_1}A_{j_2}\cdots A_{j_{m}}}$ such that
\begin{align}
\rho_{A_{j_1}A_{j_2}\cdots A_{j_{m}}}=\sum_{k}q_k\ket{\phi_k}_{A_{j_1}A_{j_2}\cdots A_{j_{m}}}\bra{\phi_k},
\end{align}
$\ket{\phi_k}_{A_{j_1}A_{j_2}\cdots A_{j_{m}}}$ is a superposition of an $m$-qudit generalized $W$-class state and vacuum.

Since each GWV state is a Schmidt rank 2 pure state under any partition, we see that for any pure state decomposition $\{p_i,\ket{\phi_i}_{A_{j_1}A_{j_2}\cdots A_{j_i}|A_{j_{i+1}}\cdots A_{j_m}}\}$ of a reduced density matrix $\rho_{A_{j_1}A_{j_2}\cdots A_{j_m}} $, $\ket{\phi_i}_{A_{j_1}A_{j_2}\cdots A_{j_i}|A_{j_{i+1}}\cdots A_{j_m}}$ is a pure state with Schmidt rank 2. We have the following theorem.

\noindent{[\bf Theorem 1]}. Let $\rho_{A_{j_1}\cdots A_{j_m}}$ be a reduced density matrix of an $n$-qudit  GW state given by (\ref{GW}). We have
\begin{align}\label{thm1}
U_{q,s}^{2}(\rho_{A_{j_1}|A_{j_2}\cdots A_{j_m}})=g_{q,s}^{2}(C^2(\rho_{A_{j_1}|A_{j_2}\cdots A_{j_m}}))
\end{align}
for $(q,s)\in\mathcal{R}$.

\begin{proof}
For convenience, we denote $\rho_{A_{j_1}|A_{j_2}\cdots A_{j_m}}$ as $\rho_{AB}$. There exists a pure
state decomposition $\{q_{i},\ket{\phi_i}\}$ such that
\begin{align}\nonumber
C(\rho_{AB})=\sum\limits_{i}q_{i}C(\ket{\phi_i}_{AB})
\end{align}
in which all $C(\ket{\phi_i}_{AB})$ are equal \cite{SX2020}. Hence, we have
\begin{align}\label{xiao}
g_{q,s}^{2}(C^{2}(\rho_{AB}))=& f_{q,s}^{2}(C(\rho_{AB}))\nonumber\\
=&f_{q,s}^{2}\left(\sum\limits_{i}q_{i}C(\ket{\phi_i}_{AB})\right)\nonumber\\
=&\sum\limits_{i}q_{i}f_{q,s}^{2}\left(C(\ket{\phi_i}_{AB})\right)\nonumber\\
=&\sum\limits_{i}q_{i}U_{q,s}^{2}\left(C(\ket{\phi_i}_{AB})\right)\nonumber\\
\geq& U_{q,s}^{2}(\rho_{AB}),
\end{align}
where the third equality is due to that $C(\ket{\phi_i}_{AB})$ are equal for all $i$,
the fourth equality is due to (\ref{D1}) and the last inequality
holds by the definition of UE.

Next we prove that $g_{q,s}^{2}(C^{2}(\rho_{AB}))\leq U_{q,s}^{2}(\rho_{AB})$. Let $\{p_{i},\ket{\omega_i}\}$
be the optimal pure state decomposition of $U_{q,s}(\rho_{AB})$. Then
\begin{align}\label{da}
U_{q,s}^{2}(\rho_{AB})=& [\sum\limits_{i}p_{i}U_{q,s}(\ket{\omega_i}_{AB})]^{2}\nonumber\\
=&[\sum\limits_{i}p_{i}f_{q,s}\left(C(\ket{\omega_i}_{AB})\right)]^{2}\nonumber\\
\geq&[f_{q,s}(\sum\limits_{i}p_{i}C(\ket{\omega_i}_{AB}))]^{2}\nonumber\\
\geq& f_{q,s}^{2}(C(\rho_{AB}))\nonumber\\
=& g_{q,s}^{2}(C^{2}(\rho_{AB})),
\end{align}
where the second equality is due to (\ref{D1}), the third inequality is due to the fact that $f_{q,s}$ is a convex function of concurrence for $(q,s)\in\mathcal{R}$ in Lemma 1, and the fourth inequality is by the definition of concurrence and the monotonicity of $f_{q,s}$.
Inequalities (\ref{xiao}) and (\ref{da}) give rise to (\ref{thm1}).
\end{proof}

The following lemma will be used to derive the monogamy relation of UE.

\noindent{[\bf Lemma 4]}. \cite{JB2008}
For any $n$-qudit GW states (\ref{GW}) and a partition $P=\{P_1,\ldots,P_r \}$
of the subsystems $S=\{A_{1}, A_2,\ldots,A_{n} \}$, $r\leq n$,
$C^{2}({\rho_{P_s|P_1\cdots\widehat{P}_s\cdots P_r}})
=\sum_{k \neq s}C^{2}(\rho_{P_s P_k})=\sum_{k \neq s}[C^{a}(\rho_{P_s P_k})]^2$
and $C(\rho_{P_s P_k})=C^{a}(\rho_{P_s P_k})$ for all $k \neq s$, where $(P_1\cdots\widehat{P}_s\cdots P_r)$ denotes that the partite $P_s$ is removed from the partition, $P_{s}\cap P_{k}=\varnothing$ for $s\neq k$ and $\bigcup_{s}P_{s}=S$.

\noindent{[\bf Theorem 2]}. Let $\rho_{A_{j_1}A_{j_2}\cdots A_{j_m}}$ be the reduced density matrix of an $n$-qudit GW state (\ref{GW}) and $\{P_1,P_2,\cdots,P_r\}$ a partition of $\{A_{j_1},A_{j_2},\cdots,A_{j_m}\},$  $r\leq m\leq n$. We have the following monogamy inequality,
\begin{align}\label{m1}
U_{q,s}^2(\rho_{P_1|P_2\cdots P_r})\geq \sum_{i=2}^{r} U_{q,s}^2(\rho_{P_1P_i})
\end{align}
for $(q,s)\in\mathcal{R}$.

\begin{proof}
For $(q,s)\in\mathcal{R}$, we have
\begin{align}
	U_{q,s}^2(\rho_{P_1|P_2\cdots P_k})=&g_{q,s}^{2}(C^2(\rho_{P_1|P_2\cdots P_r}))\nonumber\\
	=& g_{q,s}^{2}(\sum_{i=2}^{r}C^2(\rho_{P_1P_i}))\nonumber\\
	\geq & \sum_{i=2}^{r}g_{q,s}^{2}(C^2(\rho_{P_1P_i}))\nonumber\\
= & \sum_{i=2}^{r}U_{q,s}^{2}(\rho_{P_1P_i}),
\end{align}
where the first equality is due to Theorem 1, the second equality is by Lemma 4, the inequality is due to that the function $g_{q,s}^{2}(\mathcal{C}^{2})$  is a convex function in Lemma 2 for $(q,s)\in\mathcal{R}$, and the last equality is obtained by Theorem 1.
\end{proof}

When $s$ tends to 1, (\ref{m1}) is reduced to the monogamy inequality of Tsallis-$q$ entanglement given in \cite{SX2020},
${\mathcal T}_{q}^2(\rho_{P_1|P_2\cdots P_r})\geq \sum_{i=2}^{r} {\mathcal T}_{q}^2(\rho_{P_1P_i})$.When $s$ tends to 0, (\ref{m1}) reduces to the monogamy inequality of R\'enyi-$\alpha$ entanglement, ${\mathcal R}_{\alpha}^2(\rho_{P_1|P_2\cdots P_r})\geq \sum_{i=2}^{r} {\mathcal R}_{\alpha}^2(\rho_{P_1P_i})$. When $q$ tends to 1, (\ref{m1}) reduces to the monogamy inequality of EoF, $E_{f}^2(\rho_{P_1|P_2\cdots P_r})\geq \sum_{i=2}^{r} E_{f}^2(\rho_{P_1P_i})$.

Next we generalize Theorem 2 to the $\alpha$-th power of UE for GW states for $\alpha\geq2$ and $\alpha\leq0$. For $r=3$, we can always have $ U_{q,s}^{2}(\rho_{P_1P_3})\leq  U_{q,s}^{2}(\rho_{P_1P_2})$ by relabeling the partition $\{P_1,P_2,P_3\}$. Therefore, when $\alpha\geq2$ we get
\begin{eqnarray*}
  U_{q,s}^{\alpha}(\rho_{P_1|P_2P_3})
  && \geq (U_{q,s}^2(\rho_{P_1P_2})+U_{q,s}^2(\rho_{P_1P_3}))^{\frac{\alpha}{2}}\\
  &&=U_{q,s}^\alpha(\rho_{P_1|P_2})\left(1+\frac{U_{q,s}^2(\rho_{P_1P_3})}{U_{q,s}^2(\rho_{P_1P_2})}\right)^{\frac{\alpha}{2}}\\
  && \geq
  U_{q,s}^\alpha(\rho_{P_1P_2})+U_{q,s}^\alpha(\rho_{P_1P_3}),
\end{eqnarray*}
where we have used Theorem 2 in the first inequality. The second inequality is obtained since $(1+t)^x\geq 1+t^x$ for any real numbers $x$ and $t$, $0 \leq t  \leq 1$ and $x\in [1, \infty]$. When $\alpha\leq0$  we get
\begin{eqnarray*}
  U_{q,s}^{\alpha}(\rho_{P_1|P_2P_3})
  && \leq (U_{q,s}^2(\rho_{P_1P_2})+U_{q,s}^2(\rho_{P_1P_3}))^{\frac{\alpha}{2}}\\
  &&=U_{q,s}^\alpha(\rho_{P_1|P_2})\left(1+\frac{U_{q,s}^2(\rho_{P_1P_3})}{U_{q,s}^2(\rho_{P_1P_2})}\right)^{\frac{\alpha}{2}}\\
  && <
  U_{q,s}^\alpha(\rho_{P_1P_2})+U_{q,s}^\alpha(\rho_{P_1P_3}),
\end{eqnarray*}
where the first inequality is due to Theorem 2. The second inequality is obtained as $(1+t)^x< 1+t^x$ for any real numbers $x$ and $t$, $t\geq0$ and $x\in [ -\infty,0]$.
Therefore, we can have the following conclusion.

\noindent{[\bf Theorem 3]}. Let $\rho_{A_{j_1}A_{j_2}\cdots A_{j_m}}$ be the reduced density matrix of an $n$-qudit GW state $\ket{\psi}_{A_1\cdots A_n}$, and $\{P_1,P_2,\cdots,P_r\}$ a partition of $\{A_{j_1},A_{j_2},\cdots,A_{j_m}\}$, $r\leq m\leq n$. For $(q,s)\in\mathcal{R}$ we have
\begin{align}\label{m2}
U_{q,s}^\alpha(\rho_{P_1|P_2\cdots P_r})\geq \sum_{i=2}^{r} U_{q,s}^\alpha(\rho_{P_1P_i}),
\end{align}
when $\alpha\geq2$, and
\begin{align}\label{m3}
U_{q,s}^\alpha(\rho_{P_1|P_2\cdots P_r})< \sum_{i=2}^{r} U_{q,s}^\alpha(\rho_{P_1P_i}),
\end{align}
when $\alpha\leq 0$.

\section{Tighter monogamy inequalities in terms of UE}\label{SEC4}
The refined monogamy relations yield more precise characterizations of entanglement distributions, which are intimately connected to the security of quantum cryptographic  protocols~\cite{MP2010} based on entanglement. Therefore, gaining tighter entanglement monogamy relations is essential for a comprehensive grasp of quantum entanglement. Here, in term of the $\alpha$th power of UE, we provide new class of tighter monogamy relations for $n$-qudit GW states. We need the the lemma below.

\noindent{[\bf Lemma 5]}.
For any real numbers $ x\geq h\geq0$, $1\leq p\leq 1+\frac{1}{x}$ and   $m\geq 1$, we have
\begin{equation}\label{T}
\begin{aligned}
(1+x)^{m}-p^{m-1}x^{m}\geq(1+h)^{m}-p^{m-1}h^{m}.
\end{aligned}
\end{equation}

\begin{proof}
Consider the function  $f(x,m)=(1+x)^{m}-p^{m-1}x^{m}$ with $x\geq 0$, $1\leq p\leq 1+\frac{1}{x}$ and $m\geq1$. Since $\frac{\partial f(x,m)}{\partial x}=m(1+x)^{m-1}-mp^{m-1}x^{m-1}=m[(1+x)^{m-1}-(px)^{m-1}]\geq0$, the function $f(x,m)$ increases with $x$. As $x\geq h\geq 0$, we have $f(x,m)\geq f(h,m)=(1+h)^{m}-p^{m-1}h^{m}$. Therefore, we get the inequality (\ref{T}).
\end{proof}

For any tripartite state $\rho_{P_{1}P_{2}P_{3}}$, from (\ref{m1}) we have
$U_{q,s}^{2}(\rho_{P_{1}|P_{2}P_{3}})\geq U_{q,s}^{2}(\rho_{P_{1}P_{2}})+U_{q,s}^{2}(\rho_{P_{1}P_{3}})$. Therefore, there exists $\mu\ge 1$ such that
\begin{equation}\label{u}
U_{q,s}^{2}(\rho_{P_{1}|P_{2}P_{3}})\geq U_{q,s}^{2}(\rho_{P_{1}P_{2}})+\mu U_{q,s}^{2}(\rho_{P_{1}P_{3}}).
\end{equation}
By using Lemma 5 we improve the monogamy inequality (\ref{m2}) for the $\alpha$th power of UE.

\noindent{[\bf Theorem 4]}. Let $\mu\geq1$ and $h\geq1$  be any real numbers. Let $\rho_{A_{j_1}A_{j_2}\cdots A_{j_m}}$ be the reduced density matrix of an $n$-qudit GW state $\ket{\psi}_{A_1\cdots A_n}$ and $\{P_1,P_2,P_3\}$ a partition of $\{A_{j_1},A_{j_2},\cdots,A_{j_m}\},$  $3\leq m\leq n$. If $U_{q,s}^{2}(\rho_{P_{1}P_{2}})\geq h U_{q,s}^{2}(\rho_{P_{1}P_{3}})$ and $1\leq p\leq1+\frac{\mu U_{q,s}^{2}(\rho_{P_{1}P_{3}})}{U_{q,s}^{2}(\rho_{P_{1}P_{2}})}$, we have
\small\begin{align}\label{x1}
&U_{q,s}^{\alpha}(\rho_{P_{1}|P_{2}P_{3}}) \nonumber\\&\geq p^{\frac{\alpha}{2}-1}U_{q,s}^{\alpha}(\rho_{P_{1}P_{2}})+((\mu+h)^\frac{\alpha}{2}
-p^{\frac{\alpha}{2}-1}h^\frac{\alpha}{2})U_{q,s}^{\alpha}(\rho_{P_{1}P_{3}})
\end{align}
with $(q,s)\in\mathcal{R}$ and $\alpha\geq2$.\normalsize

\begin{proof}
By straightforward calculation, we have
\small\begin{align*}
&U_{q,s}^{\alpha}(\rho_{P_{1}|P_{2}P_{3}}) \nonumber \\
&=(U_{q,s}^{2}(\rho_{P_{1}|P_{2}P_{3}}))^{\frac{\alpha}{2}} \nonumber\\
&\geq(U_{q,s}^{2}(\rho_{P_{1}P_{2}})+\mu U_{q,s}^{2}(\rho_{P_{1}P_{3}}))^{\frac{\alpha}{2}}\nonumber\\
&=\mu^\frac{\alpha}{2}U_{q,s}^{\alpha}(\rho_{P_{1}P_{3}})
(1+\frac{U_{q,s}^{2}(\rho_{P_{1}P_{2}})}{\mu U_{q,s}^{2}(\rho_{P_{1}P_{3}})})^{\frac{\alpha}{2}}\nonumber\\
\end{align*}
\begin{align*}
&\geq\mu^\frac{\alpha}{2}U_{q,s}^{\alpha}(\rho_{P_{1}P_{3}})\times\nonumber\\
&~~(\frac{p^{\frac{\alpha}{2}-1}U_{q,s}^{\alpha}(\rho_{P_{1}P_{2}})}{\mu^{\frac{\alpha}{2}} U_{q,s}^{\alpha}(\rho_{P_{1}P_{3}})}+(1+\frac{h}{\mu})^{\frac{\alpha}{2}}
-p^{\frac{\alpha}{2}-1}(\frac{h}{\mu})^{\frac{\alpha}{2}})\nonumber\\
&=p^{\frac{\alpha}{2}-1}U_{q,s}^{\alpha}(\rho_{P_{1}P_{2}})+((\mu+h)^\frac{\alpha}{2}
-p^{\frac{\alpha}{2}-1}h^\frac{\alpha}{2})U_{q,s}^{\alpha}(\rho_{P_{1}P_{3}}),
\end{align*}\normalsize
where the first inequality is due to (\ref{u}) and the second inequality due to Lemma 5. Moreover, if $U_{q,s}(\rho_{P_{1}P_{2}})=0$, then $U_{q,s}(\rho_{P_{1}P_{3}})=0$ and the lower bound becomes trivially zero.
\end{proof}

\noindent{[\bf Remark 1]}. In Ref. \cite{LB2024}, the authors provide the following monogamy relation for an $n$-qudit GW state based on UE,
\small\begin{align}\label{li}
&U_{q,s}^{\alpha}(\rho_{P_{1}|P_{2}P_{3}}) \nonumber\\&\geq U_{q,s}^{\alpha}(\rho_{P_{1}P_{2}})+((\mu+h)^\frac{\alpha}{\gamma}
-h^\frac{\alpha}{\gamma})U_{q,s}^{\alpha}(\rho_{P_{1}P_{3}})
\end{align}
for $\alpha\geq\gamma$, $\gamma\geq1$, $\mu\geq1$ and $h\geq1$, with $q\geq2$, $0 \leq s \leq1$ and $qs \leq 3$. When $p=1$ and $\gamma=2$, it is obvious that inequality (\ref{li}) in Ref. \cite{LB2024} is just a special case of our Theorem 3. Moreover, it can be seen that our lower bound of the $\alpha$th power of UE becomes tighter when $p$ increases,
\small\begin{align*}
&U_{q,s}^{\alpha}(\rho_{P_{1}|P_{2}P_{3}})\\ \nonumber&\geq p^{\frac{\alpha}{2}-1}U_{q,s}^{\alpha}(\rho_{P_{1}P_{2}})+((\mu+h)^\frac{\alpha}{2}
-p^{\frac{\alpha}{2}-1}h^\frac{\alpha}{2})U_{q,s}^{\alpha}(\rho_{P_{1}P_{3}})\\ \nonumber
&\geq U_{q,s}^{\alpha}(\rho_{P_{1}P_{2}})+((\mu+h)^\frac{\alpha}{2}
-h^\frac{\alpha}{2})U_{q,s}^{\alpha}(\rho_{P_{1}P_{3}}) \nonumber
\end{align*}
for all $\alpha\geq2$, $q\geq2$, $0 \leq s \leq1$ and $qs \leq 3$, where the second equality holds when $p=1$. Hence our result (\ref{x1}) is tighter than the result (\ref{li}) given in Ref. \cite{LB2024}, see the example below.

\begin{figure}[h]
	\centering
	\scalebox{2.0}{\includegraphics[width=3.9cm]{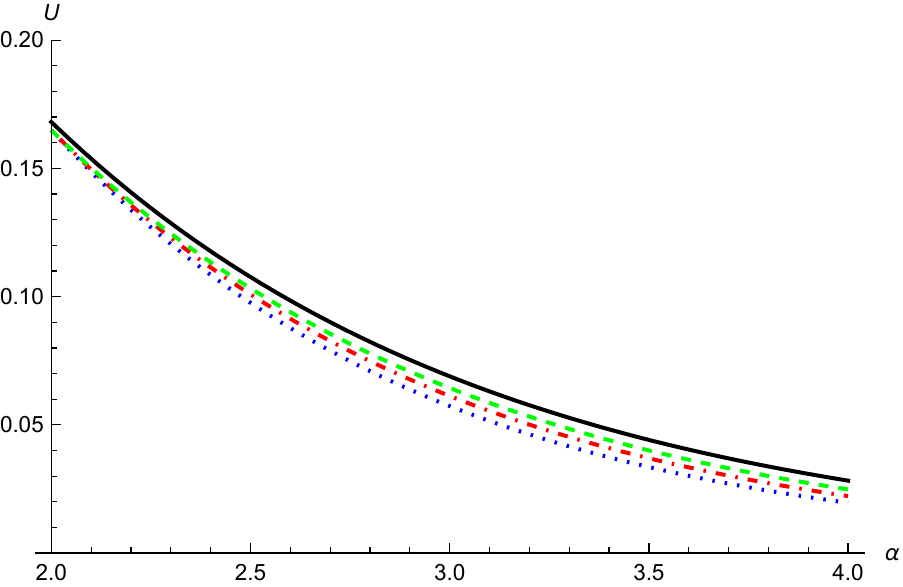}}
	\caption{\scriptsize From top to bottom, the black line is the exact values of  $U_{2,1}(\rho_{P_1|P_2P_3})$. The  green dashed line (red dotdashed line) represents the lower bound from our result (\ref{x1}) when $p=2.6$ ($p=1.8$). The blue dotted line represents the lower bound from the result (\ref{li}) in Ref. \cite{LB2024}.}
	\label{Fig1}
\end{figure}

\noindent{[\bf Example 1]}. Consider the following 4-qubit generalized $W$-class state,
\begin{align}\label{lizi1}
&\ket{\psi}_{A_1A_2A_3A_4} \nonumber \\
&=0.3\ket{0001}+0.4\ket{0010}+{0.5}\ket{0100}+\sqrt{0.5}\ket{1000}.
\end{align}
We choose $\rho_{A_1A_2A_3}$ to be the reduced density matrix of $\ket{\psi}_{A_1A_2A_3A_4}$, $P_1=A_1, P_2=A_2, P_3=A_3$. Then we have $\rho_{A_1A_2A_3}=0.09\ket{000}\bra{000}+\ket{\phi}\bra{\phi}$, where $\ket{\phi}=0.4\ket{001}+0.5\ket{010}+\sqrt{0.5}\ket{100}$. By direct calculation, we get $C(\rho_{P_1|P_2P_3})=\sqrt{\frac{41}{50}}$, $C(\rho_{P_1P_2})=\frac{\sqrt{2}}{2}$, $C(\rho_{P_1P_3})=\frac{2\sqrt{2}}{5}$. Set $\gamma=2$, $s=2$ and $q=1$. We obtain $U_{2,1}(\rho_{P_1|P_2P_3})=\frac{41}{100}$, $U_{2,1}(\rho_{P_1P_2})=\frac{1}{4}$ and $U_{2,1}(\rho_{P_1P_3})=\frac{4}{25}$. Then we can get
\small\begin{align}\label{bijiao1}
&U_{q,s}^{\alpha}(\rho_{P_{1}|P_{2}P_{3}}) \nonumber\\&\geq p^{\frac{\alpha}{2}-1}(\frac{1}{4})^{\alpha}+((\mu+h)^\frac{\alpha}{2}
-p^{\frac{\alpha}{2}-1}h^\frac{\alpha}{2})(\frac{4}{25})^{\alpha}
\end{align}
from our result (\ref{x1}) in Theorem 4, and
\small\begin{align}\label{bijiao2}
&U_{q,s}^{\alpha}(\rho_{P_{1}|P_{2}P_{3}}) \nonumber\\&\geq (\frac{1}{4})^{\alpha}+((\mu+h)^\frac{\alpha}{2}
-h^\frac{\alpha}{2})(\frac{4}{25})^{\alpha}
\end{align}
from the result (\ref{li}) given in Ref. \cite{LB2024}. Set $\mu=4$ and $h=1$. Fig.\ref{Fig1} shows that (\ref{bijiao1}) is tighter than (\ref{bijiao2}).

Note that the third system $P_{3}$ in Theorem 3 can be divided into two subsystems. Consequently, by iterately using Theorem 4 we can extend the monogamy inequality to multipartite qudit systems.

\noindent{[\bf Theorem 5]}. Let $\mu_t\geq1$ and $h_t\geq1$ be real numbers, $1\le t\le r-2$. Let $\rho_{A_{j_1}A_{j_2}\cdots A_{j_m}}$ be the reduced density matrix of an $n$-qudit GW state $\ket{\psi}_{A_1\cdots A_n}$ and $\{A,B_1,\cdots,B_{r-1}\}$ a partition of  $\{A_{j_1},A_{j_2},\cdots,A_{j_m}\},$  $r\leq m\leq n$. If $U_{q,s}^{2}(\rho_{AB_i})\geq h_{i}U_{q,s}^{2}(\rho_{A|B_{i+1}\cdots B_{r-1}})$,
$U_{q,s}^{2}(\rho_{A|B_i\cdots B_{r-1}})\geq U_{q,s}^{2}(\rho_{AB_i})+\mu_{i}U_{q,s}^{2}(\rho_{A|B_{i+1}\cdots B_{r-1}})$, $1\leq p_{i}\leq1+\frac{\mu_{i}U_{q,s}^{2}(\rho_{A|B_{i+1}\cdots B_{r-1}})}{U_{q,s}^{2}(\rho_{AB_{i}})}$ for $i=1,2,\cdots,k$,
and $U_{q,s}^{2}(\rho_{A|B_{j+1}\cdots B_{r-1}})\geq h_{j}U_{q,s}^{2}(\rho_{AB_j})$,
$U_{q,s}^{2}(\rho_{A|B_j\cdots B_{r-1}})\geq\mu_{j}U_{q,s}^{2}(\rho_{AB_j})+U_{q,s}^{2}(\rho_{A|B_{j+1}\cdots B_{r-1}})$, $1\leq p_{j}\leq\frac{\mu_{j} U_{q,s}^{2}(\rho_{AB_{j}})}{U_{q,s}^{2}(\rho_{A|B_{j+1}\cdots B_{r-1}})}$ for $j=k+1,\cdots,r-2$, $1\leq k\leq r-3$ and $r\geq4$,  then the UE satisfies
\small\begin{eqnarray}
&&U_{q,s}^{\alpha}(\rho_{P_1|P_2\cdots P_{r}})\nonumber\\
&&\geq  p^{\frac{\alpha}{2}-1}_{1}U_{q,s}^{\alpha}(\rho_{AB_1})+
\sum\limits_{i=2}^{k}\prod\limits_{l=1}^{i-1}\Gamma_{l}p^{\frac{\alpha}{2}-1}_{i}U_{q,s}^{\alpha}(\rho_{AB_{i}})\nonumber\\
&&~~~+ \Gamma_1\cdots
\Gamma_{k+1}U_{q,s}^{\alpha}(\rho_{AB_{k+1}})\nonumber\\
&&~~~+\Gamma_1\cdots\Gamma_{k}\sum\limits_{j=k+2}^{r-2}\prod\limits_{l=k+1}^{j-1}p^{\frac{\alpha}{2}-1}_{l}\Gamma_{j}U_{q,s}^{\alpha}(\rho_{AB_{j}})\nonumber\\
&&~~~+ \Gamma_1\cdots\Gamma_{k}(p_{k+1}\cdots p_{r-2})^{\frac{\alpha}{2}-1}U_{q,s}^{\alpha}(\rho_{AB_{r-1}})
\end{eqnarray}\normalsize
for all $\alpha\geq2$, where $\Gamma_{t}=(\mu_{t}+h_{t})^\frac{\alpha}{2}
-p_{t}^{\frac{\alpha}{2}-1}h_{t}^\frac{\alpha}{2}$ with $(q,s)\in\mathcal{R}$.

\begin{proof}
From Theorem 4, $U_{q,s}^{2}(\rho_{AB_i})\geq h_{i}U_{q,s}^{2}(\rho_{A|B_{i+1}\cdots B_{r-1}})$,
$U_{q,s}^{2}(\rho_{A|B_i\cdots B_{r-1}})\geq U_{q,s}^{2}(\rho_{AB_i})+\mu_{i}U_{q,s}^{2}(\rho_{A|B_{i+1}\cdots B_{r-1}})$, $1\leq p_{i}\leq1+\frac{\mu_{i}U_{q,s}^{2}(\rho_{A|B_{i+1}\cdots B_{r-1}})}{U_{q,s}^{2}(\rho_{AB_{i}})}$ for $i=1,2,\cdots,k$. We have
\begin{eqnarray}\label{x}
&&U_{q,s}^{\alpha}(\rho_{A|B_1\cdots B_{r-1}}) \nonumber\\
&&\geq p^{\frac{\alpha}{2}-1}_{1}U_{q,s}^{\alpha}(\rho_{AB_1})
+\Gamma_1U_{q,s}^{\alpha}(\rho_{A|B_2\cdots B_{r-1}})\nonumber\\
&&\geq p^{\frac{\alpha}{2}-1}_{1}U_{q,s}^{\alpha}(\rho_{AB_1})
+\Gamma_1p^{\frac{\alpha}{2}-1}_{2}U_{q,s}^{\alpha}(\rho_{AB_2})\nonumber\\
& &\quad+\Gamma_1\Gamma_{2}U_{q,s}^{\alpha}(\rho_{A|B_3\cdots B_{r-1}})\nonumber\\
&&\geq\cdots\nonumber\\
&&\geq p^{\frac{\alpha}{2}-1}_{1}U_{q,s}^{\alpha}(\rho_{AB_1})
+\Gamma_1p^{\frac{\alpha}{2}-1}_{2}U_{q,s}^{\alpha}(\rho_{AB_2})\nonumber\\
& &\quad+\cdots+\Gamma_1\cdots\Gamma_{k-1}p^{\frac{\alpha}{2}-1}_{k}U_{q,s}^{\alpha}
(\rho_{AB_k})\nonumber\\
& &\quad+\Gamma_1\cdots\Gamma_{k}U_{q,s}^{\alpha}(\rho_{A|B_{k+1}\cdots B_{r-1}}).
\end{eqnarray}
If $U_{q,s}^{2}(\rho_{A|B_{j+1}\cdots B_{r-1}})\geq h_{j}U_{q,s}^{2}(\rho_{AB_j})$,
$U_{q,s}^{2}(\rho_{A|B_j\cdots B_{r-1}})\geq\mu_{j}U_{q,s}^{2}(\rho_{AB_j})+U_{q,s}^{2}(\rho_{A|B_{j+1}\cdots B_{r-1}})$, $1\leq p_{j}\leq\frac{\mu_{j} U_{q,s}^{2}(\rho_{AB_{j}})}{U_{q,s}^{2}(\rho_{A|B_{j+1}\cdots B_{r-1}})}$ for $j=k+1,\cdots,r-2$, we get
\begin{eqnarray}\label{y}
&&U_{q,s}^{\alpha}(\rho_{A|B_{k+1}\cdots B_{r-1}})\nonumber\\
&&\geq\Gamma_{k+1}U_{q,s}^{\alpha}(\rho_{AB_{k+1}})+p^{\frac{\alpha}{2}-1}_{k+1}U_{q,s}^{\alpha}(\rho_{A|B_{k+2}\cdots B_{r-1}})\nonumber\\
&&\geq\Gamma_{k+1}U_{q,s}^{\alpha}(\rho_{AB_{k+1}})+p^{\frac{\alpha}{2}-1}_{k+1}\Gamma_{k+2}U_{q,s}^{\alpha}(\rho_{AB_{k+2}})\nonumber\\
& &\quad +(p_{k+1}p_{k+2})^{\frac{\alpha}{2}-1}U_{q,s}^{\alpha}(\rho_{A|B_{k+3}\cdots B_{r-1}})\nonumber\\
&&\geq\cdots\nonumber\\
&&\geq\Gamma_{k+1}U_{q,s}^{\alpha}(\rho_{AB_{k+1}})+p^{\frac{\alpha}{2}-1}_{k+1}\Gamma_{k+2}U_{q,s}^{\alpha}(\rho_{AB_{k+2}})\nonumber\\
& &\quad+\cdots+(p_{k+1}\cdots p_{r-3})^{\frac{\alpha}{2}-1}\Gamma_{r-2}U_{q,s}^{\alpha}(\rho_{AB_{r-2}})\nonumber\\
& &\quad+(p_{k+1}\cdots p_{r-2})^{\frac{\alpha}{2}-1}U_{q,s}^{\alpha}(\rho_{AB_{r-1}}).
\end{eqnarray}
Combining (\ref{x}) and (\ref{y}), we complete the proof.
\end{proof}

\section{Generalized monogamy relation and upper bound  for $n$-qudit systems}\label{SEC5}
In this section, for $n$-qudit systems $ABC_1...C_{n-2}$, we consider the generalized monogamy relation and upper bound satisfied by the $\beta$th ($0\leq\beta\leq1$) power of UE of an $n$-qudit GW state under the partition $AB$ and $C_1...C_{N-2}$. Before showing the results, we need the following lemmas.

\noindent{[\bf Lemma 6]}.~\cite{IJTP2019,QIP2020} For arbitrary two real numbers $x,~y$ such that $x\geq y\geq0$, one has
 $(x-y)^\beta\geq x^\beta-y^\beta$ and $(x+y)^\beta\leq x^\beta+y^\beta$ for $0\leq\beta\leq 1$.

\noindent{[\bf Lemma 7]}.~\cite{YF2017} The function $f_{q,s}(x)$ given in Eq.(\ref{D1}) satisfies $f_{q,s}(\sqrt{x^{2}+y^{2}})=f_{q,s}(x)+f_{q,s}(y)$ for $q=2$ and $\frac{1}{2}\leq s\leq1$.

\noindent{[\bf Lemma 8]}.~\cite{LB2024} Let $\rho_{A_{j_1}\cdots A_{j_m}}$ be a reduced density matrix of an $n$-qudit GW state (\ref{GW}). Then
$U_{q,s}(\rho_{A_{j_1}|A_{j_2}\cdots A_{j_m}})=f_{q,s}(C(\rho_{A_{j_1}|A_{j_2}\cdots A_{j_m}}))$ with $q\geq1$, $0\leq s\leq1$ and $qs\leq3$.

\noindent{[\bf Theorem 6]}. For an $n$-qudit GW state $\ket{\psi}_{ABC_1C_2\cdots C_{n-2}}$, the following inequality holds:
\scriptsize\begin{eqnarray}\nonumber
&&U_{q,s}^{\beta}(|\psi\rangle_{AB|C_1C_2\cdots C_{n-2}})\\ \nonumber
&&\geq\Big|\left(\sum_{i=1}^{n-2}U_{q,s}(C(\rho_{AC_i}))+U_{q,s}(C(\rho_{AB}))\right)^\beta\nonumber\\
&&~~~-\left(\sum_{i=1}^{n-2}U_{q,s}(C(\rho_{BC_i}))+U_{q,s}(C(\rho_{AB}))\right)^\beta\Big|, \\ \nonumber
\end{eqnarray}
\normalsize
where $q=2$ and $\frac{1}{2}\leq s\leq1$.

\begin{proof}
For $q=2$ and $\frac{1}{2}\leq s\leq1$ we have
\begin{align}\label{l1}
&U_{q,s}(|\psi\rangle_{A|BC_1C_2\cdots C_{n-2}})\nonumber \\
=&f_{q,s}(C(|\psi\rangle_{A|BC_1C_2\cdots C_{n-2}}))\nonumber \\
=&f_{q,s}(\sqrt{C^{2}(\rho_{AB})+C^{2}(\rho_{AC_{1}})+\cdots+C^{2}(\rho_{AC_{n-2}})})\nonumber \\
=& \sum_{i=1}^{n-2}f_{q,s}(C(\rho_{AC_i}))+f_{q,s}(C(\rho_{AB}))\nonumber \\
=& \sum_{i=1}^{n-2}U_{q,s}(C(\rho_{AC_i}))+U_{q,s}(C(\rho_{AB})),
\end{align}
where the first and fourth equalities are due to Lemma 8, the second and third equalities are due to Lemma 4 and Lemma 7, respectively. Similarly, we have
\begin{align}\label{l2}
&U_{q,s}(|\psi\rangle_{B|AC_1C_2\cdots C_{n-2}})\nonumber \\
=& \sum_{i=1}^{n-2}U_{q,s}(C(\rho_{BC_i}))+U_{q,s}(C(\rho_{AB})).
\end{align}

From the subadditivity of the unified-($q,s$) entropy for a quantum state $\rho_{AB}$ \cite{AE2011}, $|U_{q,s}(\rho_{A})-U_{q,s}(\rho_{B})| \leq U_{q,s}(\rho_{AB})\leq U_{q,s}(\rho_{A})+U_{q,s}(\rho_{B})$, we have
\begin{align}
&U_{q,s}^{\beta}(|\psi\rangle_{AB|C_1C_2\cdots C_{n-2}})=U_{q,s}^{\beta}(\rho_{AB})\nonumber \\
\geq&|U_{q,s}(\rho_{A})-U_{q,s}(\rho_{B})|^{\beta}\nonumber \\
\geq&|U_{q,s}^{\beta}(\rho_{A})-U_{q,s}^\beta(\rho_{B})|\nonumber\\
=&\Big|\left(\sum_{i=1}^{n-2}U_{q,s}(C(\rho_{AC_i}))+U_{q,s}(C(\rho_{AB}))\right)^\beta\nonumber\\
&-\left(\sum_{i=1}^{n-2}U_{q,s}(C(\rho_{BC_i}))+U_{q,s}(C(\rho_{AB}))\right)^\beta\Big|,\nonumber
\end{align}
where the second inequality is due to the subadditivity
of UE, the third inequality is by Lemma 6, the last equality is obtained from Eqs. (\ref{l1}) and (\ref{l2}).
\end{proof}

\noindent{[\bf Remark 2]}. In Ref. \cite{LB2024} the authors give the following generalized monogamy relation,
\begin{align*}
&U_{q,s}(|\psi\rangle_{AB|C_1C_2\cdots C_{n-2}})\\
\geq&\Big|\sum_{i=1}^{n-2}[U_{q,s}(C(\rho_{AC_i}))-U_{q,s}(C(\rho_{BC_i})]\Big|,
\end{align*}
which is obviously a special case of our Theorem 6 when $\beta=1$. Moreover, when $\beta=1$ and $s = 1$, Theorem 5 reduces to the generalized monogamy relation in terms of the Tsallis-2 entanglement given in Ref. \cite{SX2020}.

According to the subadditivity of the unified-($q,s$) entropy, we also have the following upper bound of UE.

\noindent{[\bf Theorem 7]}. For an $n$-qudit GW state $\ket{\psi}_{ABC_1C_2\cdots C_{n-2}}$, the following inequality holds,
\scriptsize\begin{eqnarray}\nonumber
&&U_{q,s}^{\beta}(|\psi\rangle_{AB|C_1C_2\cdots C_{n-2}})\\ \nonumber
&&\leq\left(\sum_{i=1}^{n-2}U_{q,s}(C(\rho_{AC_i}))+U_{q,s}(C(\rho_{AB}))\right)^\beta\nonumber\\
&&~~~+\left(\sum_{i=1}^{n-2}U_{q,s}(C(\rho_{BC_i}))+U_{q,s}(C(\rho_{AB}))\right)^\beta, \\ \nonumber
\end{eqnarray}
\normalsize
where $q=2$ and $\frac{1}{2}\leq s\leq1$.

\begin{proof}
Using the subadditivity of unified-($q,s$) entropy, we have the following inequality,
\begin{align*}
&U_{q,s}^{\beta}(|\psi\rangle_{AB|C_1C_2\cdots C_{n-2}})=U_{q,s}^{\beta}(\rho_{AB})\nonumber \\
\leq&(U_{q,s}(\rho_{A})+U_{q,s}(\rho_{B}))^{\beta}\nonumber \\
\leq&U_{q,s}^{\beta}(\rho_{A})+U_{q,s}^\beta(\rho_{B})\nonumber\\
=&\left(\sum_{i=1}^{n-2}U_{q,s}(C(\rho_{AC_i}))+U_{q,s}(C(\rho_{AB}))\right)^\beta\nonumber\\
&+\left(\sum_{i=1}^{n-2}U_{q,s}(C(\rho_{BC_i}))+U_{q,s}(C(\rho_{AB}))\right)^\beta, \\ \nonumber
\end{align*}
where the second inequality is due to the subadditivity of UE, the third inequality is by Lemma 6, the last equality is obtained by using Eqs. (\ref{l1}) and (\ref{l2}).
\end{proof}

\section{Applications}\label{SEC6}
To investigate the entanglement properties of multi-qubit $W$-class states, in this section we first present two monogamy-like inequalities of PREs for GW states by utilizing Theorem 2. For an $n$-qubit GW state $\ket\varphi_{A_{1}A_{2}\cdots A_{n}}$, we can always divide the whole system into two subsystems under any partition $P=\{P_{1}, P_{2}\}$ of $\{A_{1}, \cdots, A_{n}\}$ such that $P_{1}=\{A_{1}, A_{2}, \cdots, A_{m}\}$ and  $P_{2}=\{A_{m+1}, A_{m+2}, \cdots, A_{n}\}$. Each subsystem can be further partitioned as $P_{1}=\{P_{11}, P_{12}\}$ and $P_{2}=\{P_{21}, P_{22}\}$, where $P_{11}=\{A_{1},  \cdots, A_{a}\}$, $P_{12}=\{A_{a},  \cdots, A_{m}\}$ and $P_{21}=\{A_{m+1},  \cdots, A_{b}\}$, $P_{22}=\{A_{b+1},\cdots, A_{n}\}$. Since the partition $P=\{P_{1}, P_{2}, \cdots, P_{r}\}$ in Theorem 2 is arbitrary, if we take $r=3$ and $P=\{P_{11}P_{12}, P_{21}, P_{22}\}$, according to the monogamy relation in Theorem 2 we obtain
\begin{align}\label{A1}
&U_{q,s}^{2}(\rho_{P_{11}P_{12}|P_{21}P_{22}})\nonumber \\
\geq &U_{q,s}^{2}(\rho_{P_{11}P_{12}|P_{21}})+U_{q,s}^{2}(\rho_{P_{11}P_{12}|P_{22}})\nonumber\\
\geq &U_{q,s}^{2}(\rho_{P_{11}|P_{21}})+U_{q,s}^{2}(\rho_{P_{12}|P_{21}})\nonumber\\
 &+U_{q,s}^{2}(\rho_{P_{11}|P_{22}})+U_{q,s}^{2}(\rho_{P_{12}|P_{22}})\nonumber\\
\geq&\sum\limits_{i=1}^{m}\sum\limits_{j=m+1}^{n}U_{q,s}^{2}(\rho_{A_{i}A_{j}})
\end{align}
by repeatedly applying the second inequality. Clearly, the above inequality relations hold for any partition given by $a, m$ and $b$ with $1 \leq a \leq m <b \leq n$.

From inequality (\ref{A1}), we have the following PREs in terms of the unified-($q,s$) entanglement,
\begin{align}\label{A2}
\Upsilon_{P_{11}P_{12}|P_{21}P_{22}}
=&U_{q,s}^{2}(\rho_{P_{11}P_{12}|P_{21}P_{22}})-U_{q,s}^{2}(\rho_{P_{11}|P_{21}})\nonumber\\
&-U_{q,s}^{2}(\rho_{P_{12}|P_{21}})-U_{q,s}^{2}(\rho_{P_{11}|P_{22}})\nonumber\\
&-U_{q,s}^{2}(\rho_{P_{12}|P_{22}})
\end{align}
and
\begin{align}\label{A3}
\Upsilon'_{P_{11}P_{12}|P_{21}P_{22}}
=&U_{q,s}^{2}(\rho_{P_{11}P_{12}|P_{21}P_{22}})\nonumber\\
&-\sum\limits_{i=1}^{m}\sum\limits_{j=m+1}^{n}U_{q,s}^{2}(\rho_{A_{i}A_{j}}).
\end{align}

The monogamy inequality (\ref{A1}) and the PREs (\ref{A2}) and (\ref{A3}) contain all possible bipartitions for $n$-qubit systems. The PRE captures entanglement properties under arbitrary partitions, elucidating various multi-partitioned entanglements with respect to $a,m$ and $b$. Let us take the $n$-qubit $W$ state $\ket W_{A_{1}A_{2}\cdots A_{n}}$ to illustrate our inequalities and PREs,
$$
\ket W_{A_{1}A_{2}\cdots A_{n}}=\frac{1}{\sqrt N}(|10\cdots0\rangle+|01\cdots0\rangle+|0\cdots01\rangle).
$$
For $(q,s)\in\mathcal{R}$, using Theorem 1 we obtain
\begin{align}\label{A4}
\Upsilon_{P_{11}P_{12}|P_{21}P_{22}}
=&g_{q,s}^{2}(C_{P_{11}P_{12}|P_{21}P_{22}})-g_{q,s}^{2}(C_{P_{11}|P_{21}})\nonumber\\
&-g_{q,s}^{2}(C_{P_{12}|P_{21}})-g_{q,s}^{2}(C_{P_{11}|P_{22}})\nonumber\\
&-g_{q,s}^{2}(C_{P_{12}|P_{22}})
\end{align}
and
\begin{align}\label{A5}
\Upsilon'_{P_{11}P_{12}|P_{21}P_{22}}
=&g_{q,s}^{2}(C_{P_{11}P_{12}|P_{21}P_{22}})\nonumber\\
&-\sum\limits_{i=1}^{m}\sum\limits_{j=m+1}^{n}g_{q,s}^{2}(C_{A_{i}A_{j}}).
\end{align}
By direct calculation we have $C^{2}_{P_{11}P_{12}|P_{21}P_{22}}=\frac{4m(n-m)}{n^{2}}$, $C^{2}_{P_{11}P_{21}}=\frac{1}{n^{2}}[\sqrt{(n-m)^{2}+4a(b-m)}-(n-m)]^{2}$,
$C^{2}_{P_{12}P_{21}}=\frac{1}{n^{2}}[\sqrt{(n-m)^{2}+4(m-a)(b-m)}-(n-m)]^{2}$,
$C^{2}_{P_{11}P_{22}}=\frac{1}{n^{2}}[\sqrt{(n-m)^{2}+4a(n-b)}-(n-m)]^{2}$,
$C^{2}_{P_{12}P_{22}}=\frac{1}{n^{2}}[\sqrt{(n-m)^{2}+4(m-a)(n-b)}-(n-m)]^{2}$ and
$C^{2}_{A_{i}A_{j}}=\frac{1}{n^{2}}[\sqrt{4+(n-2)^{2}}-(n-2)]^{2}$ \cite{LYY2020,MZX2010}.

When $s$ tends to 1, the unified-($q,s$) entropy converges to Tsallis-$q$ entropy. In this case the PRE $\Upsilon_{P_{11}P_{12}|P_{21}P_{22}}$ (\ref{A2}) relies on the two partitions in terms of $a, m$ and $b$. Let $n=6$ and $m=4$. Then $a$ can be 1, 2, 3 and 4, and $b$ be 5 or 6. For all the possible values of $a$ and $b$ the value of the PRE $\Upsilon_{P_{11}P_{12}|P_{21}P_{22}}$ for a 6-qubit W-class state is shown in Tables (\ref{T1}) and (\ref{T2}).
\begin{table}[h!]
\centering
\begin{tabular}{||c c c c c||}
\hline
q & $\Upsilon(q,1,5)$ & $\Upsilon(q,2,5)$ & $\Upsilon(q,3,5)$&$\Upsilon(q,4,5)$ \\ [0.5ex]
\hline\hline
2.0 & 0.191172 & 0.193981 & 0.191172& 0.183117\\
2.1 & 0.179591 & 0.182176 & 0.179591 &0.172876\\
2.2 & 0.168899 & 0.171295 & 0.168899 &0.162006\\
2.3& 0.159022 & 0.161259 & 0.159022 &0.152585\\
2.4 & 0.149893 & 0.151993 & 0.149893 &0.143847\\ [1ex]
\hline
\end{tabular}
\caption{\scriptsize The values of PRE $\Upsilon_{P_{11}P_{12}|P_{21}P_{22}}$ for different $q$ and all the possible values of $a$, denoted by $\Upsilon(q,a,b)$, when $s=1, m=4, b=5, n=6$.}
		\label{T1}
	\end{table}
\begin{table}[h!]
		\centering
		\begin{tabular}{||c c c c c||}
			\hline
			q & $\Upsilon(q,1,6)$ & $\Upsilon(q,2,6)$ & $\Upsilon(q,3,6)$&$\Upsilon(q,4,6)$ \\ [0.5ex]
			\hline\hline
			2.0 & 0.173999 & 0.183117 & 0.173999& 0.148145\\
			2.1 & 0.163680 & 0.172876 & 0.163680 &0.139568\\
			2.2 & 0.154082 & 0.162006 & 0.154082 &0.131526\\
			2.3& 0.145158 & 0.152585 & 0.145158 &0.123996\\
			2.4 & 0.136863 & 0.143847 & 0.136863 &0.116952\\ [1ex]
			\hline
		\end{tabular}
		\caption{\scriptsize The values of PRE $\Upsilon_{P_{11}P_{12}|P_{21}P_{22}}$  for different $q$ and all the possible values of $a$, denoted by $\Upsilon(q,a,b)$, when $s=1, m=4, b=6, n=6$.}
		\label{T2}
	\end{table}

\begin{figure}[h]
	\centering
	\scalebox{2}{\includegraphics[width=3.9cm]{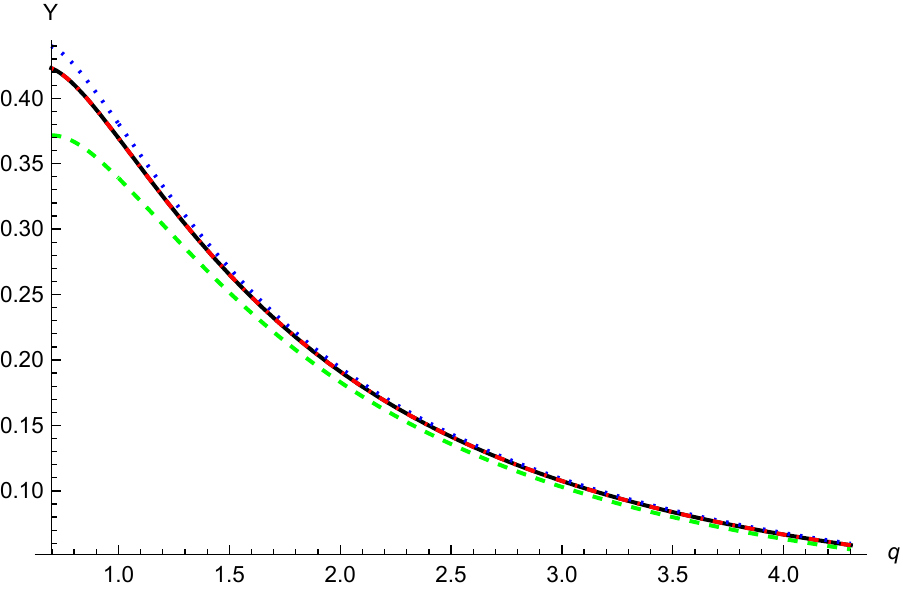}}
	\caption{\scriptsize $s=1, m=4, b=5, n=6$. From top to bottom, blue dotted line represents $a=2$, black line represents $a=1$, red dotdashed line represents $a=3$, green dashed line represents $a=4$. The curves coincide when $a=1$ and $a=3$.}
	\label{Fig2}
\end{figure}

\begin{figure}[h]
	\centering
	\scalebox{2}{\includegraphics[width=3.9cm]{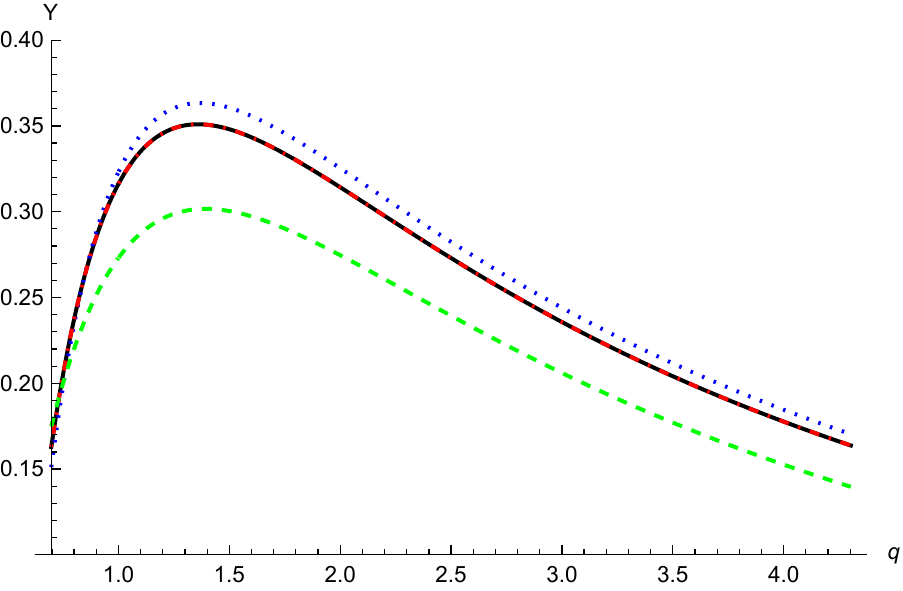}}
	\caption{\scriptsize $s=1, m=4, b=6, n=6$. From top to bottom, blue dotted line represents $a=2$, black line represents $a=1$, red dotdashed line represents $a=3$, green dashed line represents $a=4$. The curves coincide when $a=1$ and $a=3$.}
	\label{Fig3}
\end{figure}

We observe that the values in Table (\ref{T1}) with $b=5$ are greater than those in Table (\ref{T2}) with $b=6$, and the values of $\Upsilon(q, 4, 5)$ and $\Upsilon(q, 2, 6)$ are equal. In Table (\ref{T1}) (Table (\ref{T2})), the PRE has the same value when $a=1$ and $a=3$, and the PRE gets the maximum value when $a=2$ and the minimum when $a=4$. If $a, m, b$ are fixed for $\frac{5-\sqrt{13}}{2}\leq q\leq\frac{5+\sqrt{13}}{2}$, the value of PRE decreases with the increase of $q$, as shown in Fig.\ref{Fig2} and  Fig.\ref{Fig3}, corresponding to Tables (\ref{T1}) and (\ref{T2}), respectively.

On the other hand, the PRE $\Upsilon'_{P_{11}P_{12}|P_{21}P_{22}}$ in Eq. (\ref{A3}) is only related to the first partition in terms of $m$. Set $n=6$. Then the value of $m$ can be 1, 2, 3, 4, 5. The values of PRE $\Upsilon'_{P_{11}P_{12}|P_{21}P_{22}}$ are shown in Table (\ref{T3}) when $s=1$.
\begin{table}[h!]
		\centering
		\begin{tabular}{||c c c c c c||}
			\hline
			q & $\Upsilon'(q,1)$ & $\Upsilon'(q,2)$ & $\Upsilon'(q,3)$&$\Upsilon'(q,4)$&$\Upsilon'(q,5)$ \\ [0.5ex]
			\hline\hline
			2.0 & 0.077113 & 0.197450 & 0.249914& 0.197450&0.077113\\
			2.1 & 0.071820 & 0.185350 & 0.235131 &0.185350&0.071820\\
			2.2 & 0.067131 & 0.174229 & 0.221395 &0.174229&0.067131\\
			2.3& 0.062961 & 0.163993 & 0.208622 &0.163993&0.062961\\
			2.4 & 0.059234 & 0.154560 & 0.196737 &0.154560&0.059234\\ [1ex]
			\hline
		\end{tabular}
\caption{\scriptsize The values of PRE $\Upsilon'_{P_{11}P_{12}|P_{21}P_{22}}$ for different $q$ and all the possible values of $m$, denoted by $\Upsilon'(q,m)$, when $s=1$ and $n=6$.}
		\label{T3}
	\end{table}

\begin{figure}[h]
	\centering
	\scalebox{2}{\includegraphics[width=3.9cm]{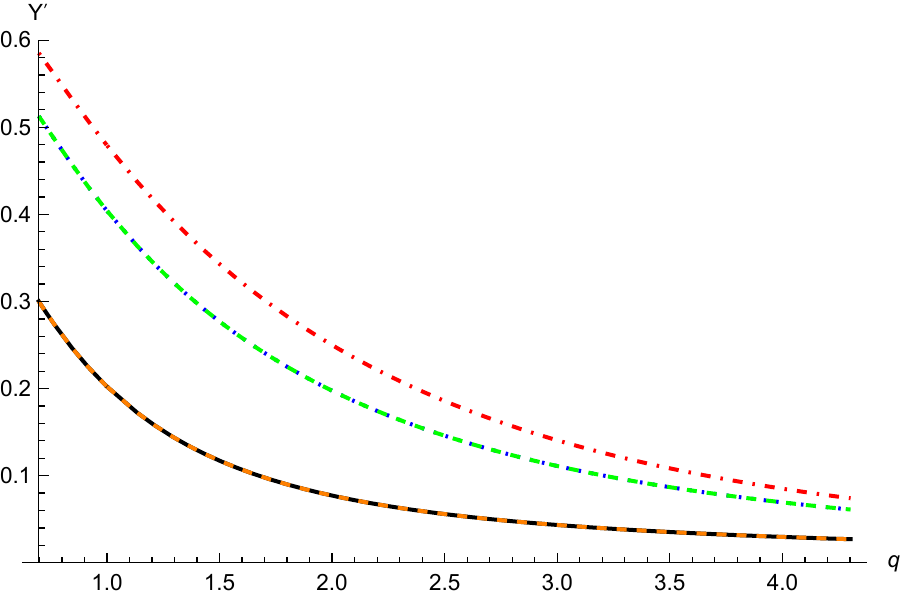}}
	\caption{\scriptsize $s=1,n=6$. From top to bottom, red dotdashed line represents $m=3$, blue dotted line represents $m=2$, green dashed line represents $m=4$, black line represents $m=1$, orange dashed line represents $m=5$. The curves coincide when $a=1$ ($a=2$) and $a=5$ ($a=4$).}
	\label{Fig4}
\end{figure}

We see that the value of PRE $\Upsilon'_{P_{11}P_{12}|P_{21}P_{22}}$ is the same when $m=1$ ($m=2$) and $m=5$ ($m=4$), and the value is maximum when $m=3$. The above results are also due to the structural particularity of the $W$-class state. If $m,n$ are fixed for $\frac{5-\sqrt{13}}{2}\leq q\leq\frac{5+\sqrt{13}}{2}$, the value of PRE $\Upsilon'_{P_{11}P_{12}|P_{21}P_{22}}$ decreases with the increase of $q$, as shown in Fig. \ref{Fig4}, corresponding to the case in Table (\ref{T3}). These results of PREs show the entanglement structures of the six-qubit $W$ state.

\section{Conclusion}\label{SEC7}
The monogamy relations of quantum entanglement are the essential characteristics displayed by multiqudit entangled states. We have investigated monogamy properties related to the unified-($q,s$) entropy for $n$-qudit GW states under any partition. We have provided an analytical formula of the unified-($q,s$) entanglement with extended $(q,s)$ region for $(q,s)\in\mathcal{R}$. By using the analytical formula, the monogamy relation based on the squared UE for a qudit GW state has been presented. Since the distribution of entanglement in multi-qudit systems can be better understood by employing stricter monogamy inequalities, we have  also investigated tighter monogamy relations based on the $\alpha$th ($\alpha\geq2$) power of the  UE. Furthermore, for $n$-qudit systems $ABC_1...C_{n-2}$, generalized monogamy relation and upper bound satisfied by the $\beta$th ($0\leq\beta\leq1$) power of UE for the GW states have been established under the partition $AB$ and $C_1...C_{n-2}$. To demonstrate the significance of our conclusions, we have presented the two partition-dependent residual entanglements to give a comprehensive analysis of the entanglement structure of the GW states. Our results indicate that the UE serves as an effective measure of entanglement in multi-qudit systems in the MoE framework. When the parameters $q$ and $s$ of the unified-($q,s$) entanglement converge to some values, our results turn to be the ones for other entanglement measures. Our results may shed new light on further investigations on comprehending the distribution of entanglement in other multipartite systems.

\bigskip
\noindent\textbf{Acknowledgments} This work is supported by the National Natural Science Foundation of China under Grants 12075159 and 12171044, the specific research fund of the Innovation Platform for Academicians of Hainan Province.

\bigskip
\noindent{\bf Data availability statement}\
No new data were created or analyzed in this study.

\end{document}